# Precise and Fast LIDAR via Electrical Asynchronous Sampling Based on a Single Femtosecond Laser

**Lizong Dong, Qinggai Mi, Siyu Zhou, Yuetang Yang, Yuanzu Wang and Guanhao Wu***

*State Key Laboratory of Precision Measurement Technology and Instruments, Department of Precision Instrument, Tsinghua University, Beijing 100084, China*

**Corresponding author: guanhaowu@mail.tsinghua.edu.cn*

**Abstract:** Using a laser-based ranging method for precise environmental 3D sensing, LiDAR has numerous applications in science and industry. However, conventional LiDAR face challenges in simultaneously achieving high ranging precision and fast measurement rates, which limits their applicability in more precise fields, such as aerospace, smart healthcare and beyond. By employing an asynchronous electrical pulse sampling strategy on a single optical frequency comb with a stable repetition rate and femtosecond-pulse width, we exploit the advantages of optical-frequency-comb ranging method and overcome the limitations of sampling aliasing and low data-utilization inherent in traditional approaches. This enables a significant improvement of LiDAR's performance to achieve micrometer-level precision and megahertz-regimes update rates over meter-range on non-cooperative targets. Specifically, we achieve 38.8-μm Allan deviation at 1-MHz update rate and 8.06-μm Allan deviation after 2-ms time-averaging based on a 56.091-MHz femtosecond laser. This enhancement enables various advanced measurement applications, including metrology monitoring on high-speed objects, 1-megapixel/s precise 3D scanning imaging and first-ever contactless vital sign detection using time-of-flight LiDAR. This LiDAR unlock new possibilities for precise and fast real-time measurements in diverse fields.

## Introduction

Light detection and ranging (LiDAR) is crucial in the intelligent perception of autonomous vehicles and robots[1-4]. Intelligent driving systems must respond timely and precisely when confronted with unforeseen events, such as the presence of children or unknown obstacles. Precise and fast LiDAR detection allows decision-making for additional time to provide timely protection for vulnerable lives in such situations. This renders it applicable in unexpected application as well[5-7]. For instance, in advanced manufacturing, it can be employed to monitor the conditions of aircraft turbine blades during high-speed rotation. In intelligent healthcare, high-resolution non-contact vital sign detection using LiDAR, such as the simultaneous monitoring of human movement, breathing, and heartbeat information, exceeds imagination. LiDAR relies on the following two ranging technological approaches: time-of-flight (TOF) and frequency-modulated continuous-wave (FMCW) methods[8-11]. The precision and speed of conventional TOF are usually in the centimeter and kilohertz regimes, respectively, limited by the light source performance and bandwidth of the electrical devices[3]. In the FMCW method, the speed, range resolution, and accuracy depend heavily on the stringent conditions of the frequency, agility, and linearity of the lasers. Achieving high precision requires the laser frequency to be both stable and tunable, which limits the system's ability to operate at higher speeds. Additionally, complex signal processing at the backend leads to a trade-off between real-time, faster speed, and higher precision, posing a significant challenge to swept-frequency light sources[10,11].

Recently, numerous studies have focused on utilizing femtosecond lasers as light sources to achieve rapid and high precision measurement[12-15]. Dual comb ranging (DCR) is one of the most effective methods. The fiber-DCR method employs asynchronous optical sampling and exhibits outstanding performance, including micrometer-level precision and meter-level non-ambiguous range (NAR). Most DCR systems measure distances using the Fast Fourier Transform (FFT). To avoid the spectral aliasing problem and enable the FFT, the update rate of the fiber dual comb is limited to the kilohertz range owing to its lower repetition frequency[16]. To overcome the constraint, electro-optic (EO) combs and microcombs have been developed to achieve ultrahigh repetition rates that increase the update rate to the megahertz level[17-19]. However, owing to the high repetition rate of EO combs or microcombs, the NAR is very small and proportional to the inverse of the repetition rate, thus restricting their application to larger-range measurements[20-25]. Real-time processing is also an important issue in high-speed measurement. The data processing and evaluation is often performed offline in most investigations due to their low data utilization rate and large data volume. For dual-comb methods, the data utilization rate (the ratio of the measurement rate to the repetition rate of the comb) is generally low whether using fiber-based comb or microcomb. For commonly used ~100-MHz repetition-rate fiber-combs, the acquisition rate is limited to several kHz for adequate sampling. For microcombs over tens of GHz regime repetition-rate, they could only reach acquisition rate in MHz regime. Moreover, the distance is calculated using the FFT, fitting or other time-consuming algorithm, which relies on massive amount of data and requires sufficient time for complex computation[15,16]. As a result, it would impose additional burden on acquisition device and computational resources. However, increasing sampling rate restricts the storage depth and is limited by the speed of the analog-to-digital converter (ADC), cost, and associated large-volume data processing requirements, necessitating the division of the acquisition and processing steps. These factors pose a significant challenge toward achieving real-time systems[20,24].

To solve these problems, we propose a simple and powerful TOF measurement method using electrical asynchronous sampling based on a single femtosecond laser. This approach effectively overcomes the trade-offs between measurement precision and speed of fiber-combs, while also enhancing the data utilization. This method enables high precision and fast update rate while maintaining a large NAR, thereby surpassing the limitations of state-of-the-art dual-comb optical asynchronous sampling methods. Owing to the higher data utilization and smaller data volume, it also possible to facilitate real-time computation and data output based on integrated devices. In contrast to the conventional applications of frequency combs, we demonstrated several novel applications of the proposed LiDAR, including noncontact vital-sign detection, fast metrology real-time monitoring, and 3D imaging.

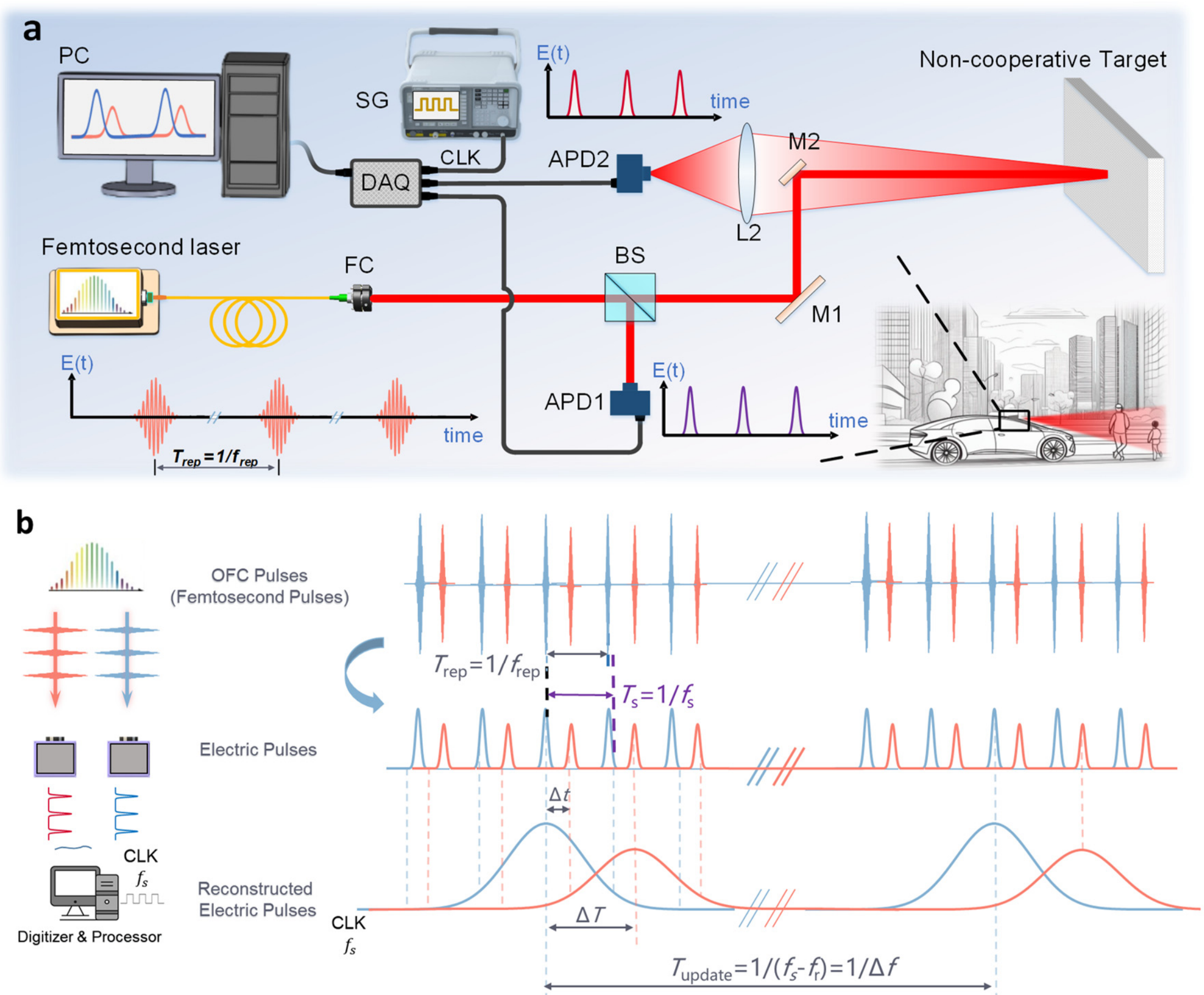

**Fig. 1. Schematic and principle of the electrical asynchronous sampling ranging method. a,** Schematic of the optical system. BS: beam splitter; FC: fiber collimator; APD1, APD2: avalanche photodetectors; M1, M2: mirrors; Femtosecond fiber-laser with stable repetition rate. **b,** Principle of the asynchronous sampling method. A pair of repetition-rate-locked femtosecond laser trains are detected and photoelectrically converted using APD, forming a pair of broadened electrical pulse trains. The waveforms are further broadened in time domain by asynchronous sampling under the sampling clock $f_s$. CLK: clock with $f_s = f_{rep} + \Delta f$ sine waveform. Optical pulses are converted to electrical pulses and high-repetition-frequency electrical pulses are reconstructed. Blue and orange pulses represent the reference and measurement signals, respectively. $T_{rep}$ and $T_s$ denote the periods of pulses and sampling, respectively.

## Principles of electrical asynchronous sampling ranging method

The proposed approach is based on a coaxial optical path, as shown in Fig. 1(a). It indirectly measures the TOF of the optical pulses by comparing the time delay between two periodic electrical waveforms from avalanche photodiodes (APD), APD1 and APD2. The measurements are obtained from the two distinct pulse trains using a femtosecond laser with a locked frequency rate $f_{\mathrm{rep}}$ divided by beam splitter (BS). The pulse train serving as the reference beam is directed to APD1 for converting the optical pulses into electrical pulses. The other pulse train, serving as the target beam, is incident on the target. The returning pulse is focused on APD2 through the emission and receiving coaxial optical paths. Two electrical pulse trains are recorded by one digitizer. The reference and measured electrical pulse signals are simultaneously digitized and recorded under a stable clock-sampling signal $f_{\mathrm{s}}$, as shown in Fig. 1(b). Subsequently, the point data are processed by combining the pulse location algorithm and system parameters. The scanner could be applied to the coaxial optical path to realize precise 3D imaging.

To process two high-repetition-rate electrical waveforms, we propose an asynchronous sampling ranging method for electrical pulses. We employ a sampling rate ($f_{\mathrm{s}} = f_{\mathrm{rep}} + \Delta f$) to the digitizer supplied by a frequency synthesizer. At the set clock rate, a single data point is captured per electrical pulse cycle. Subsequently, the next data point with a minute pulse shift is captured when the next clock cycle arrives. This process is iteratively performed before amplifying the waveforms and original time delays ($\Delta t$) of the target and reference beams, as depicted in Fig. 1(b). The two pulse trains are amplified by a factor of $f_{\mathrm{rep}}/\Delta f$ in time domain to yield a pair of reconstructed pulses under a measured period ($T_{\mathrm{update}}=1/\Delta f$), which represents the measurement acquisition rate. The relative amplified time delay ($\Delta T$) is obtained by calculating the centroid between the amplified reference and target pulses. Subsequently, combined with the amplification coefficient, light speed, and refractive index of air, the distance between the reference target and the object is derived as,

$$L = \frac{v_g}{2} \cdot \Delta T \cdot \frac{\Delta f}{f_{rep}} = \frac{v_g}{2} \cdot \Delta t \tag{1}$$

where $v_{\mathrm{g}}$ is the group velocity of the femtosecond laser in air, $\Delta T$ is the amplified time-delay, and $\Delta t$ is the original time delay. The electrical signals of the high-repetition-frequency femtosecond pulses are reconstructed using an asynchronous sampling ranging method that overcame the bandwidth limitations of conventional devices.

The principle for achieving micrometer-level precision measurements is the amplification of the original minimal-time delay through asynchronous sampling, which allows timing on a larger time scale. This time-scale amplification allows for more precise timing operations and reduction in the data volume. As a result, precise and fast distance detection can be achieved using low-cost acquisition devices and reduced computational resources. Moreover, the asynchronous sampling method can be further extended to the periodic multi-sampling method with a faster update rate and periodic under-sampling method with a lower requirement for the acquisition component (see Section 1 of Supplementary Information). The acquisition rate and precision can be switched by adjusting the sampling clock rate and selecting a matching algorithm without laser tuning. The following experimental data with slightly different setting update rates were obtained using this method.

## Results

### Precision analysis

To evaluate the precision of the proposed asynchronous sampling ranging method, the distance from a fixed noncooperative target was measured at an update rate of 1 MHz with 57.091-MHz sampling rate and 56.091-MHz laser repetition rate. The measurement target here is an unpainted cardboard. Combined with the APD's conversion gain, we measured the received optical power to be 0.52 μW. The Allan deviation, which was used to assess the repeatability of the proposed method, was analyzed for the 5-s data point, as shown in Fig. 2(a). The precision of the original data was 406.8 μm without time-averaging and it decreased to 8.06 μm with an averaging of 2 ms. Figure 2(b) shows the 5ms measurement results, including the original data (blue point) and Kalman filtered data (red point). To further suppress the noise in the original data while maintaining the measurement rate, the Karman filter (KF) was applied to improve the precision to 38.8 μm at an update rate of 1 MHz（see Section 5 of Supplementary Information for Kilman filer)[26]. For distance measurement, the accuracy of the proposed system was assessed via comparative experiments using a commercial laser interferometer (PT-313B) with an accuracy of 1 nm. A moving target corner mirror positioned at ~1 m was measured at a discrete moving step of 1 cm and range of 20 cm on the guide rail. Figures 2(c) and (d) show the measurement accuracy result of the proposed system. Both ranging results were calculated under identical environmental parameters, including temperature, humidity, and air pressure, inducing an uncertainty of ~$10^{-7}$. By applying linear fitting [Fig. 2(c)], the slope and correlation coefficient ($R^2$) were obtained as 1.00078223 and 0.999999925, respectively. The residuals were within ±50 μm with a standard deviation of 23.74 μm [Fig. 2(d)] due to the electrical noise of the detector.

The upper precision limit of proposed method is influenced by a combination of factors, including the rate difference $\Delta f$, intensity noise induced by APD and acquisition devices, pulse location algorithm, and the electrical pulse jitter, etc. (see Section 2 of Supplementary Information for detailed discussions).

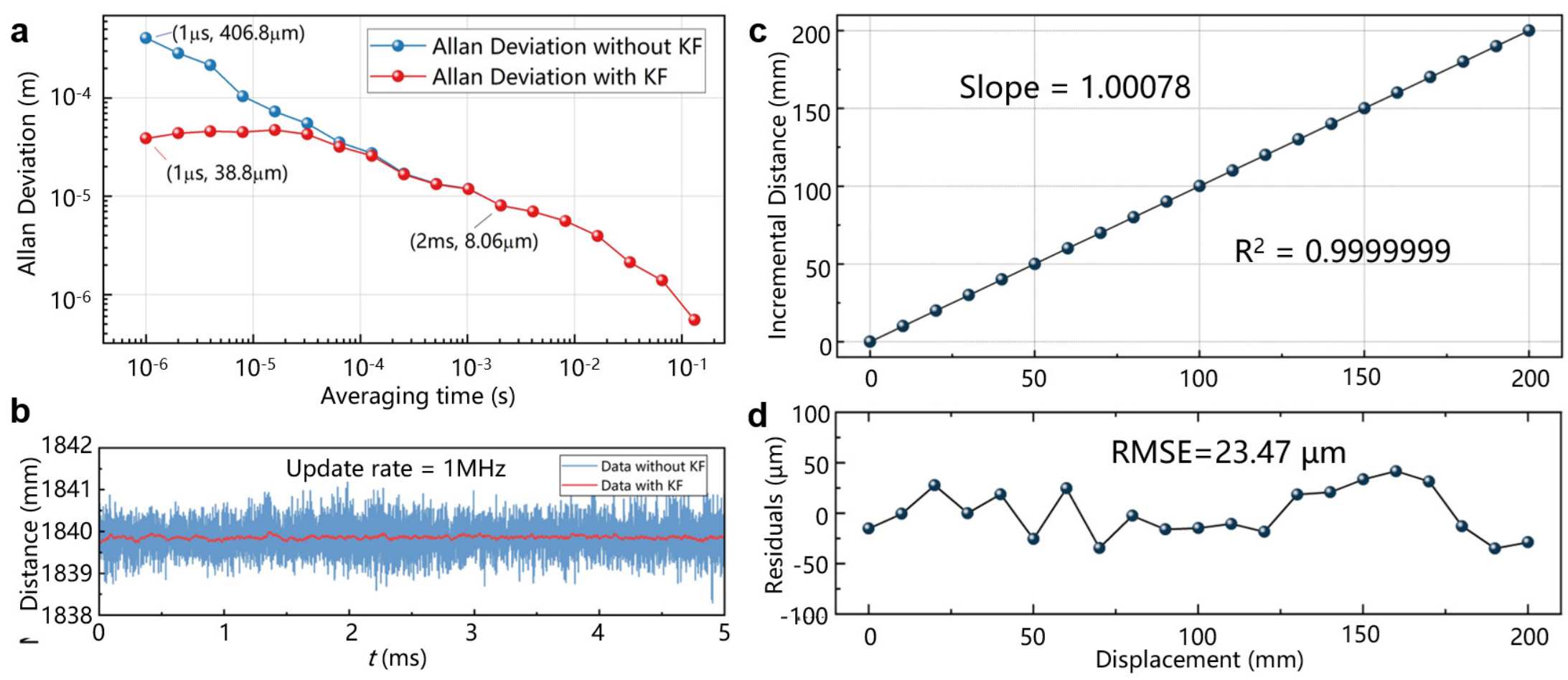

**Fig. 2. Precision analysis. a,** Allan deviation of the distance measurement versus averaging time evaluated over the 5 s data points. Blue and red lines represent the original measurement results without filter and with the Kalman filter, respectively. After averaging for 2 ms, the precision of the two measurements decreased to 8.06 μm. **b,** Results of 1 MHz TOF measurements within an acquisition time of 5 ms. **c,** Experimental evaluation of the measurement accuracy compared with that obtained using a commercial laser interferometer. Comparison of incremental distance measured by the proposed ranging system (vertical axis) with the displacements measured by a commercial laser interferometer (horizontal axis) at a distance of ~1 m with a moving step of ~10 mm. **d,** Residuals within ±50 μm.

## Verification of fast metrology monitoring

The state of turbine blades at different rotational speeds and vibration modes significantly affects the overall engine performance. Turbines often operate at speed exceeding ten thousand revolutions per minute (RPM); however, the existing laser measurement methods have not yet met the demands of real-time fast metrology monitoring (see the application scene in Fig. 3(a)). The proposed measurement method offers a novel solution for monitoring the rotation state of a turbine.

The profile of a rapidly rotating fan was measured to validate the reproducibility and capability of the proposed system for rapid metrology. In this experiment, the measurement beam was focused on the surface of the fan blade, which rotated at a speed exceeding 10000 RPM, as shown in Fig. 3(a). The drawing shows the rotating direction of the fan and fixed detection position of the laser measurement. The distance-acquisition rate was set to 500 kHz with 56.591 MHz sampling rate and 56.091MHz laser repetition rate. The returned optical power of fan profile was approximately 0.5 μW and that of the background was 1.3 μW when the emitted optical power is around 10 mW. The dynamic measurement results, including the data for the fan blade and fixed clamping element, were reconstructed as shown in Fig. 3(b). The Karman filter results (black line) were used to eliminate the noise at the cost of a long transition time. At a rotation speed of 6000 RPM, the fan blade profile was simultaneously captured and reconstructed using over 300 data points. Owing to the insufficient dynamic balance characteristics of the fan, the high rotation speed of the fan at 10000 RPM caused vibrations in a specific range affecting the supporting device. Figure 3(c) shows a sinusoidal resonance background movement with an amplitude of 2 mm, precisely matching the rotation rate of 166.6 Hz.

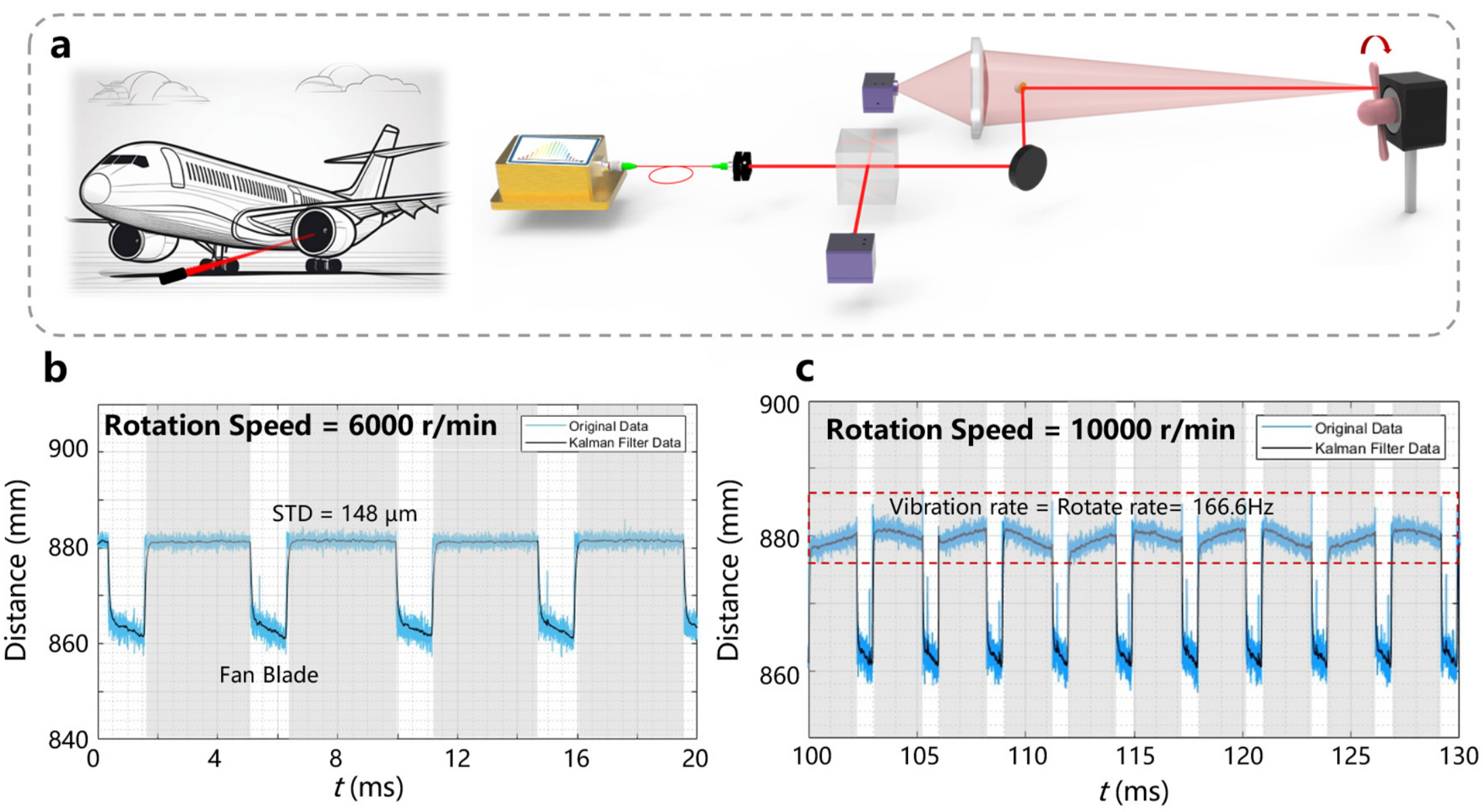

**Fig. 3. Results of fast metrology of a fast-rotating object. a,** Application to turbine monitoring and measurement for high-rotation-speed fan. The laser light is emitted by the previously mentioned experimental setup toward the fan blade. The recorded data include the distance and other status information. **b,** Measurement result at 6000 RPM. The standard deviation of the background is approximately 148 μm. **c,** Measurement result at 10000 RPM. The vibration rate of background is equal to the rotation rate.

### 3D scanning imaging

Based on the validated precise and fast measurement method, we further complete the submillimeter-level resolution and meter-level NAR 3D scanning imaging on non-cooperative targets. The acquisition rate was 1 MHz with 57.091-MHz sampling rate and 56.091-MHz laser repetition rate, i.e., the 3D imaging system attained a point rate of 1 MP/s using a single fiber laser. A scene containing a model gate of Tsinghua University, two model cars, and "THU" characters with a distance spacing in the range 5–10 cm were created and positioned at a distance of ~1 m in front of the scanning mirror [see Fig. 4(a)]. The material of model is ABS plastic (Acrylonitrile Butadiene Styrene), with a low reflectivity for 1550-nm wavelength light. Additionally, there are variations in reflected light intensity due to different colors and incident angles. The returned optical power of a white block at normal incidence angle is about 0.14μW when the emitted optical power is around 20 mW. Figure 4(b) shows the front and top views of the reconstructed scene 3D image with a resolution of 180000 pixels (600 L, 300 H, each point was calculated as the average of 50 ranging results). A portion of the detailed point cloud was represented by characters "THU" [see Fig. 4(d)]. The protrusions on "THU" are the joints of the building blocks. Each joint had a radius of 4.8 mm and height of 1.7 mm. Figure 4(e) shows the details of the joint point cloud. The excellent vertical and horizontal range resolutions of the imaging system facilitate the clear visualization of sub-millimeter-level details at the meter-level scale.

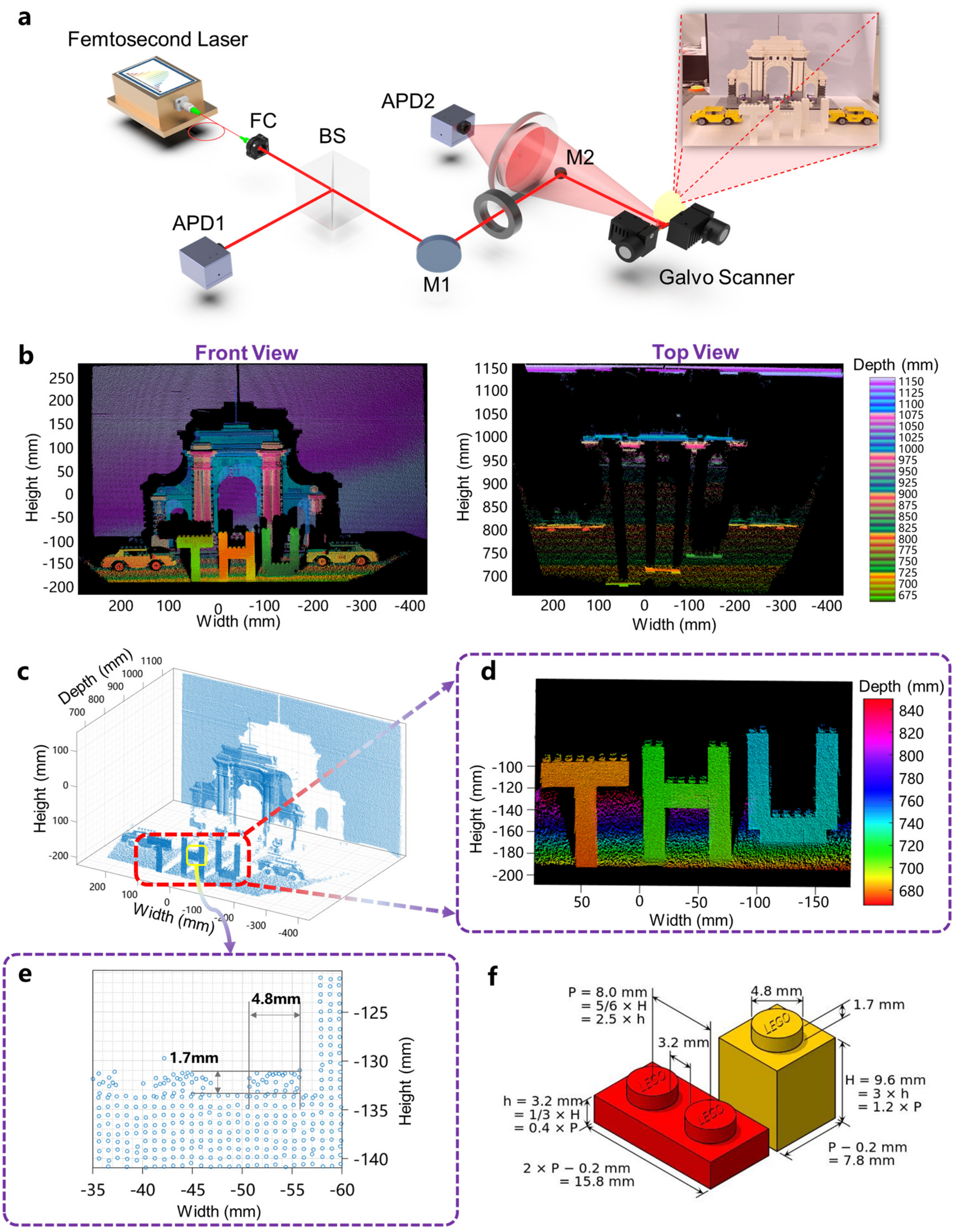

**Fig. 4. 3D scanning imaging. a,** Experimental setup. Laser beam is steered by the galvanometer scanner. The constructed scene includes various types of plastic models. **b,** Front and top views of the point cloud with linear striped pseudo colored rendering. **c,** 3D image reconstructed by the combination of distance measurement and angle-record. **d,** Details of the "THU" characters. **e,** Details of the protrusions, which are the joints in the building blocks. Each protrusion has a radius and height of 4.8 and 1.7 mm, respectively. **f,** The dimension diagram of the experimental model blocks.

## Noncontact vital sign monitoring

To further validate the excellent distance resolution and universality of the proposed LiDAR system, an experiment was performed to assess the vital signs of the human body, including breath and heartbeat information. Vital sign detection is widely implemented in medical and health settings using contact and wearable sensors. However, they are unsuitable for patients with burn wounds or other skin conditions, necessitating noncontact detection. Figure 5(a) shows the experimental setup involving femtosecond lasers directed at the chest of the participant. This approach allowed the observation of variations in chest contours caused by breathing and heartbeat movements, thereby enabling the measurement of breathing rate, heart rate, and other physiological parameters. Typically, the respiratory movement can cause micro-motions in the chest wall of approximately 1 to 12 mm, whereas the cardiac movement can generate micro-motions of approximately 100 to 500 μm. The heartbeat of healthy adults typically falls in the range 0.8–2 Hz, whereas the breathing rate is in the range 0.1–0.5 Hz. These two signals can be extracted and identified separately using different bandpass filters.

The distance-acquisition rate was set to 500 kHz with 56.591MHz sampling rate, and the laser repetition rate is 56.091MHz. By adjusting the emitted optical power to 10 mW, we quantified the received optical power reflected from the skin at 0.5 meters to be approximately 0.1 μW. Figure 5(b) shows the chest motion information obtained from the measurements on the participant while sitting. The respiratory rate is approximately 0.34 Hz (20.4 breaths per min) and the distance variation caused by the chest motion is approximately 5 mm. After removing the motion artifacts and extracting them using a bandpass filter, the heartbeat-induced chest contour fluctuations were obtained. These fluctuations exhibited an amplitude of approximately 300 μm at a frequency of 1.54 Hz (92.4 heartbeats per min), as indicated by the red line. The time–frequency graph was obtained by calculating the data in Fig. 5(b) using a short-time FFT. Three purple peaks were observed at all times, with frequencies increasing in the following order: respiratory signal, second harmonic of the respiratory signal, and heartbeat signal. The time–frequency graph obtained by performing a short-time FFT with the normalized amplitude was plotted, yielding a breathing rate and heart rate of 20.4/min and 92.4/min, respectively, as shown in Fig. 5(c). Two vibration rates can be observed in the spectrum obtained from the Fourier transform of the total signal, with the amplitude of the respiratory motion being approximately 22 dB higher than that of the heartbeat motion.

The high-range resolution of the proposed system enabled accurate respiratory detection during human breathing. Unlike traditional contact-based methods such as electrocardiograms (ECG) or photoplethysmography (PPG), which require continuous skin contact and may be uncomfortable for some patients, especially for patients with sensitive skin, burns or other dermatological conditions, the laser-based approach offers a non-contact alternative. The method's ability to capture various motion signals with high precision suggests its potential for use in other areas, such as body positioning or even surgical navigation, where non-invasive monitoring could be beneficial. This approach may offer a promising direction for the development of future intelligent healthcare solutions, potentially providing a safer and more adaptable option for patient monitoring and medical applications.

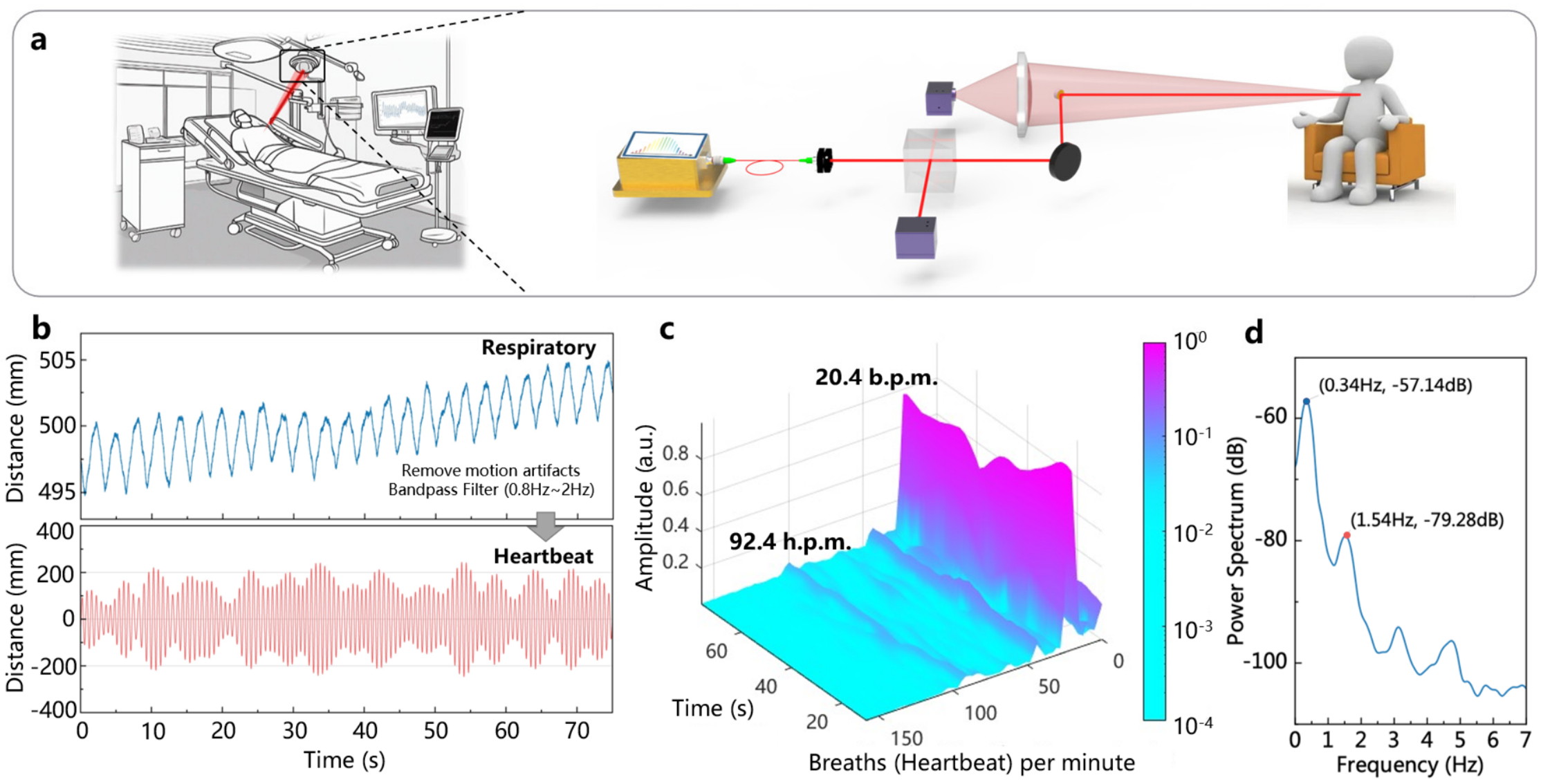

**Fig. 5. Results of single-point measurement for vital signs monitoring. a,** Schematic application of vital sign measurement and experimental setup. The laser is emitted by the proposed experimental setup and directed toward the chest area for monitoring. **b,** Results of overall signs, including breaths and heartbeat movements shown as the blue line. The heartbeat signal is obtained after specific filtering (bandwidth 0.8 to 2 Hz), shown as the red line data with an amplitude of approximately 300 μm. **c,** Time–frequency graph obtained by performing the short-time FFT on the data in (b) with a normalized amplitude normalized. The breathing and heart rates are 20.4 and 92.4 per minute, respectively. **d,** Power spectrum shows that the respiratory rate is 0.34 Hz and heartbeat rate is approximately 1.54 Hz. The amplitude of breaths is 22 dB higher than that of heartbeats.

## Conclusion

This study achieved high-precision and high-speed femtosecond LiDAR based on asynchronous sampling of electrical pulses with a single 56.091MHz repetition rate fiber-laser, yielding a precision of 38.8 μm at an update rate of 1 MHz and 8.06 μm after time-averaging of 2 ms. This overcomes the limitations of measurement rate and ranging precision of the conventional TOF LiDAR systems. Compared to state-of-the-art dual-comb methods, the proposed method overcomes the Nyquist sampling limit of interferometric optical asynchronous sampling. The excellent data utilization rate ($1.78 \times 10^{-2}$, 1MHz/56.091MHz, see Supplementary section for details) leads to simpler distance calculation and shorter computation time, providing the system with potential ultrafast real-time processing and data output capability. Experiments verified the outstanding performance of the proposed method in the fast metrology of profiles and vibrations, high-range resolution 3D imaging, and real-time detection of human vital signs. The method balanced the measurement rate, precision, and range and enhanced LiDAR for unexpected scenarios. While the results are promising, there is still room for further exploration. The proposed femtosecond LiDAR can achieve a faster measurement rate by increasing the repetition rate of the femtosecond laser and the bandwidth of detectors, including the use of an EO comb or micro-comb with a repetition frequency of up to GHz[17,27]. Higher precision, longer range, and small-volume integration of products could be achieved by transforming the spatial optical system structure into a fiber optical path, applying an Er-doped optical fiber amplifier in the receiver optical path, and changing the APD to a single-photon detector. Furthermore, low-cost compact fiber lasers and prospective on-chip femtosecond lasers can potentially reduce the cost of the LIDAR system, paving the way for intelligent applications.

# Methods

## Experiment setup

A homemade Er-doped fiber femtosecond laser is used in experiment. The laser operates at a repetition rate of 56.091 MHz, locked to an RF reference source by feedback control of the cavity length, achieving a frequency stability (Allan deviation) of $3 \times 10^{-12}$. The pulse width of femtosecond laser from FC is approximately 160 fs, measured by an autocorrelator (pulseCheck NX50, APE). In experimental setup, a 90% transmission and 10% reflection beam splitter is installed in the optical path to separate the reference and measurement light. In measurement path, we use a plano-convex lens with 60 mm focal length and 2 inches diameter for converging scattered light back from non-cooperative target. Two InGaAs APDs (KY-APRM-XX-I-FS) with a bandwidth of 1GHz are employed for the optical-to-electrical pulse conversion. For data acquisition, a dual-channel 16-bit digitizer capable of receiving an external clock (10MHz-150MHz) is used to simultaneously collect data at this sampling rate. After the APD generates an analog signal, it is converted by an Analog-to-Digital Converter (ADC) and stored in a First-In, First-Out (FIFO) hardware device. The data are subsequently transferred to a computer via Peripheral Component Interconnect Express (PCI-E) interface using Direct Memory Access (DMA) for processing and evaluation.

In precision verification of Section 2.2, a commercial laser interferometer (PT-313B, with a resolution up to 1 nm) serves as the ground truth estimator. The target consists of two back-to-back corner cubes mounted on a single-axis motorized stage (OSMS-CS-26-200, Optosigma) for synchronously movement.

In the experiment detailed in Section 2.4, a Galvo scanner (QS10XY-AG, Thorlabs) is incorporated into the coaxial optical path to facilitate 3D imaging. For scenarios that require measuring local details, a focusing lens can be used in the optical path to converge parallel emission beams, achieving higher lateral resolution, see optical path in Fig.4(a). The triggers for the Galvo scanning driver and data collection are set to operate synchronously. The measured distance, along with the scanning angle readings, are converted into Cartesian coordinates, resulting in a 3D point cloud. Both point-by-point scanning and continuous-scanning can be implemented by programming the Galvo scanner.

## Data processing and evaluation

MATLAB calculation program captures the transmitted data in real-time, performs distance computation and realizes 3D-reconstruction. We use the centroid method for distance computation. The reconstructed waveform of asynchronous sampling consists of a series of discrete points. Due to the electrical characteristics of the APD, resulting in slight asymmetry in the rising and falling edges but approximating a Gaussian distribution. As shown in Figure 1(b), the blue points represent the reconstructed reference pulses, while the red points represent the reconstructed measurement pulses. The pulse broadening is attributed to the limited bandwidth of APD, which can be considered as passing through a low-pass filter, leading to a loss of high-frequency optical information. Given the asymmetry of reconstructed waveform and the time-consuming figures of the fitting method, we utilize the centroid method to calculate the flight time delay of the reconstructed pulses. Therefore, the broadening effect of the electrical pulse assists in the pulse timing.

The formula for calculating the pulse centroid is structured as a weighted average of the sampling times ($t_n$) of each point, where the weights are the corresponding voltage amplitudes ($U_n$) of the reconstructed electrical pulses. The formula for calculating the arrival time of the pulse is as follows,

$$t_{centroid} = \frac{\sum_{i=1}^{n} U_i t_i}{\sum_{i=1}^{n} U_i} \tag{2}$$

For a pair of reconstructed pulses at 1MHz acquisition rate, excluding non-pulse data, only 30 data points (16-bit each) are involved in the centroid method calculation. This simplification of the algorithm reduces computation time, making it feasible for real-time implementation on an Field-Programmable Gate Array (FPGA) in future work.

To reduce the impact of the random process on distance measurement, a one-dimensional discrete dynamic model of the proposed system based on the Kalman filter was used. It achieves effective noise suppression for distance measurements of both static and dynamic targets, as demonstrated in section 2.2 and 2.3, respectively. More details for Kalman filter method are given in Section 5 of Supplementary Information.

For 3D imaging, we establish the mapping relationship between the rotation angle of scanner, calculated distance, and the Cartesian coordinate System. In the experiment, real-time 3D reconstruction is achieved by synchronously recording the rotation angle of the mirror and the real-time distance evaluation.

**Availability of data and materials**

Data underlying the results presented in this paper are not publicly available at this time but may be obtained from the authors upon reasonable request.

**Conflict of interest**

The authors declare that they have no competing interests.

**Funding**

National Natural Science Foundation of China (Grants No. 62227822, 52327805, 52105555, 92150104)

**Authors' contributions**

G.W. led this work. L.D. wrote the software for experiments and data processing. L.D., Q.M. and S.Z. performed the measurement experiments. Y.Y. optimized the light source. L.D., S.Z., Y.W., Y.Y. and G.W. prepared the manuscript. All authors discussed the results and contributed to the manuscript.

**Acknowledgments**

We thank Prof. Yang Li and Prof. Yuanmu Yang for their helpful suggestions and discussion. We thank Hao Ouyang and Xingyu Jia for their help with figures and experiments.

**Supplemental information**

See Supplementary 1 for supporting content.

**References**

1. Schwarz, B. Mapping the world in 3D. *Nat. Photonics* **4**, 429–430 (2010).
2. Royo, S. & Ballesta-Garcia, M. An overview of lidar imaging systems for autonomous vehicles. *Appl. Sci.* **9**, 4093 (2019).
3. Behroozpour, B., Sandborn, P. A. M., Wu, M. C. & Boser, B. E. Lidar system architectures and circuits. *IEEE Commun. Mag.* **55**, 135–142 (2017).
4. Roriz, R., Cabral, J. & Gomes, T. Automotive LiDAR technology: A survey. *IEEE Trans. Intell. Transp. Syst.* **23**, 6282–6297 (2021).
5. Zhang, Z., Liu, Y., Stephens, T. & Eggleton, B. J. Photonic radar for contactless vital sign detection. *Nat. Photonics* **17**, 791–797 (2023).
6. Dong, L., Mi, Q., Zhou, S. *et al.* High-speed and high-precision ranging system for non-cooperative targets based on optical frequency comb. *CLEO: Applications and Technology*. Optica Publishing Group, **7**, ATh1K (2023).
7. Teleanu, E. L. & Durán, V. Electro-optic dual-comb interferometer for high-speed vibrometry. *Opt. Express* **25**, 16427–16436 (2017).
8. Kuse, N. & Fermann, M. E. Frequency-modulated comb LIDAR. *APL Photonics* **4**, 100 (2019).
9. Lukashchuk, A., Riemensberger, J., Karpov, M. *et al.* Dual chirped microcomb based parallel ranging at megapixel-line rates. *Nat. Commun.* **13**, 3280 (2022).
10. Roos, P. A., Reibel, R. R., Berg, T. *et al.* Ultrabroadband optical chirp linearization for precision metrology applications. *Opt. Lett.* **34**, 3692–3694 (2009).
11. Komissarov, R., Kozlov, V., Filonov, D. *et al.* Partially coherent radar unties range resolution from bandwidth limitations. *Nat. Commun.* **10**, 1423 (2019).
12. Coddington, I., Swann, W. C., Nenadovic, L. & Newbury, N. R. Rapid and precise absolute distance measurements at long range. *Nat. Photonics* **3**, 351–356 (2009).
13. Lee, J., Kim, Y. J., Lee, K., Lee, S. & Kim, S. Time-of-flight measurement with femtosecond light pulses. *Nat. Photonics* **4**, 716–720 (2010).
14. Wang, J., Lu, Z., Wang, W. *et al.* Long-distance ranging with high precision using a soliton microcomb. *Photon. Res.* **8**, 1964–1972 (2020).
15. Zhu, Z. & Wu, G. Dual-comb ranging. *Engineering* **4**, 772–778 (2018).
16. Jiang, R., Zhou, S. & Wu, G. Aliasing-free dual-comb ranging system based on free-running fiber lasers. *Opt. Express* **29**, 33527–33535 (2021).
17. Zhuang, R., Ni, K., Wu, G. *et al.* Electro-optic frequency combs: Theory, characteristics, and applications. *Laser Photon. Rev.* **17**, 2200353 (2023).
18. Zhang, M., Buscaino, B., Wang, C. *et al.* Broadband electro-optic frequency comb generation in a lithium niobate microring resonator. *Nature* **568**, 373–377 (2019).
19. Suh, M. G. & Vahala, K. J. Soliton microcomb range measurement. *Science* **359**, 884–887 (2018).
20. Na, Y., Jeon, C. G., Ahn, C. *et al.* Ultrafast, sub-nanometre-precision and multifunctional time-of-flight detection. *Nat. Photonics* **14**, 355–360 (2020).
21. Zhang, X., Kwon, K., Henriksson, J. *et al.* A large-scale microelectromechanical-systems-based silicon photonics LiDAR. *Nature* **603**, 253–258 (2022).
22. Chen, R., Shu, H., Shen, B. *et al.* Breaking the temporal and frequency congestion of LiDAR by parallel chaos. *Nat. Photonics* **17**, 306–314 (2023).
23. Riemensberger, J., Lukashchuk, A., Karpov, M. *et al.* Massively parallel coherent laser ranging using a soliton microcomb. *Nature* **581**, 164–170 (2020).
24. Trocha, P., Karpov, M., Ganin, D. *et al.* Ultrafast optical ranging using microresonator soliton frequency combs. *Science* **359**, 887–891 (2018).
25. Bao, C., Suh, M. G. & Vahala, K. Microresonator soliton dual-comb imaging. *Optica* **6**, 1110–1116 (2019).

26. Hu, D., Wu, Z., Cao, H. *et al.* Dual-comb absolute distance measurement of non-cooperative targets with a single free-running mode-locked fiber laser. *Opt. Commun.* **482**, 126566 (2021).
27. Snigirev, V., Riedhauser, A., Lihachev, G. *et al.* Ultrafast tunable lasers using lithium niobate integrated photonics. *Nature* **615**, 411–417 (2023).